\documentclass[a4paper,11pt]{article}
\usepackage{pos}

\title{The muon anomalous magnetic moment: is the lattice spacing small enough?}
\ShortTitle{Muon anomalous magnetic moment}

\author[a]{Christopher Aubin}
\author[b]{Thomas Blum}
\author*[c]{Maarten Golterman}
\author[d]{Santiago Peris}

\affiliation[a]{Department of Physics \& Engineering Physics Fordham University, Bronx,
New York, NY 10458, USA}
\affiliation[b]{Physics Department
University of Connecticut, Storrs, CT 06269, USA}
\affiliation[c]{Department of Physics and Astronomy, San Francisco State University
San Francisco, CA 94132, USA}
\affiliation[d]{Department of Physics and IFAE-BIST, Universitat Aut\`onoma de Barcelona
E-08193 Bellaterra, Barcelona, Spain}

\emailAdd{maarten@sfsu.edu}

\abstract{We present new results for the light-quark connected part of the leading order
hadronic-vacuum-polarization (HVP) contribution to the muon anomalous magnetic moment,
using $2+1+1$ staggered fermions. We have collected more statistics on previous ensembles, and
we added two new ensembles. This allows us to reduce statistical errors on the HVP
contribution and related window quantities significantly. We also calculated the
current-current correlator to next-to-next-to-leading order (NNLO) in staggered chiral
perturbation theory, so that we can correct to NNLO for finite-volume, pion-mass
mistuning and taste-breaking effects. We discuss the applicability of NNLO chiral
perturbation theory, emphasizing that it provides a systematic EFT approach to the HVP
contribution, but not to short- or intermediate-distance window quantities.  This makes
it difficult to assess systematic errors on the standard intermediate-distance window
quantity that is now widely considered in the literature. In view of this, we
investigate a longer-distance window, for which EFT methods should be more reliable.
Our most important conclusion is that new high-statistics computations at lattice
spacings significantly smaller than 0.06 fm are indispensable.  The ensembles we use
have been generously provided by MILC and CalLat.}

\FullConference{%
The 39th International Symposium on Lattice Field Theory,\\
8th-13th August, 2022,\\
Rheinische Friedrich-Wilhelms-Universität Bonn, Bonn, Germany
}

\begin{document}
\maketitle
	
\section{Introduction}

In this talk, we review our recent work on the light-quark connected contribution to the hadronic
vacuum polarization (HVP) contribution to the muon anomalous magnetic momentum,
$a_\mu^{\rm HVP}$ \cite{Aubin:2022hgm}.
We begin with a brief discussion of the applicability of chiral perturbation theory (ChPT) to $a_\mu^{\rm HVP}$,
reviewing also our previous work in Ref.~\cite{Aubin:2020scy}, after which we present our new
staggered
results for the light-quark connected part of $a_\mu^{\rm HVP}$ and the ``standard'' intermediate window
quantity of Ref.~\cite{RBC18}, as well as a new window quantity introduced in Ref.~\cite{Aubin:2022hgm}.   Our main focus is on the continuum limit.

We summarize our main conclusions:
\begin{itemize}
\item NNLO (next-to-next-to-leading-order) ChPT works for $a_\mu^{\rm HVP}$ if all pions
(including heavier taste pions) are light enough, with masses less than $\lesssim 280$~MeV.
(This corrects a statement in Ref.~\cite{Golterman:2017njs}.)   It does not work for the standard
intermediate window, for which no effective-field-theory (EFT) is available.
\item To control the continuum limit with staggered fermions, lattice spacings smaller than 0.06~fm
will be needed.
\end{itemize}
Our computations were done using HISQ ensembles with lattice spacings 0.06, 0.09, 0.12, and 0.15~fm provided by the MILC collaboration
\cite{MILC:2012znn}, as well as a larger-volume 0.15~fm ensemble provided by the CalLat collaboration \cite{Miller:2020xhy} ({\it cf.} Table~\ref{tab1} below).

\begin{boldmath}
\section{$a_\mu^{\rm HVP}$ and chiral perturbation theory}
\end{boldmath}
$a_\mu^{\rm HVP}$ is a low-energy quantity, involving muons, photons, pions, and other hadrons.
It is thus described by an EFT for the lowest-mass particles among these, pions coupled to muons
and photons, with the effects of heavier hadrons represented by low-energy constants (LECs) in this
EFT, which is constructed by coupling ChPT to photons and muons.   

$a_\mu^{\rm HVP}$ can be written in terms of an integral over euclidean momentum as
\begin{equation}
\label{amu}
a_\mu^{\rm HVP}=4\alpha^2\int_0^\infty dQ^2\,f(Q^2)\,\hat\Pi(Q^2)\ ,
\end{equation}
where $\alpha$ is the fine-structure constant, $\hat\Pi(Q^2)$ is the subtracted scalar HVP, and $f(Q^2)\sim m_\mu^4/Q^6$ (for large $Q^2$) is a 
known function \cite{LR,TB}.   In ChPT, at order N$^k$LO, modulo logarithms,
\begin{equation}
\label{hatPi}
\hat\Pi(Q^2)\sim (Q^2)^{k-1}
\end{equation}
for large $Q^2$.
This implies that to NNLO the integral over $Q^2$ in Eq.~(\ref{amu}) is finite, and the counter terms needed
are just those of ChPT.   Beyond NNLO, the integral diverges, and new counter terms in the extended EFT including also photons and muons are needed.   Reference~\cite{Aubin:2020scy} gives the relevant counter term at N$^3$LO as
\begin{equation}
\label{ct}
\frac{\alpha^2 m_\mu^3}{(4\pi f_\pi)^4}\,(\bar\mu\sigma_{\alpha\beta}F_{\alpha\beta}\mu)\,\mbox{tr}(Q\Sigma Q\Sigma^\dagger)\ ,
\end{equation}
with $\Sigma$ the non-linear pion field, $\mu$ the muon field, and $Q=\mbox{diag}\left(\frac{2}{3},-\frac{1}{3}\right)$ the quark fractional charge matrix.   

At NNLO, we can thus systematically compute $a_\mu^{\rm HVP}$ in ChPT.   Using the known values of the 
muon and pion masses, $f_\pi$, the one-loop
counter term $\ell_6$ and the two-loop counter term $c_{56}$, we find that $a_\mu^{\rm HVP}=660(160)\times 10^{-10}$, with most of the error coming from the uncertainty in $c_{56}$ \cite{Aubin:2022hgm}.

\begin{figure}[t!]
\vspace*{4ex}
\begin{center}
\includegraphics*[width=9cm]{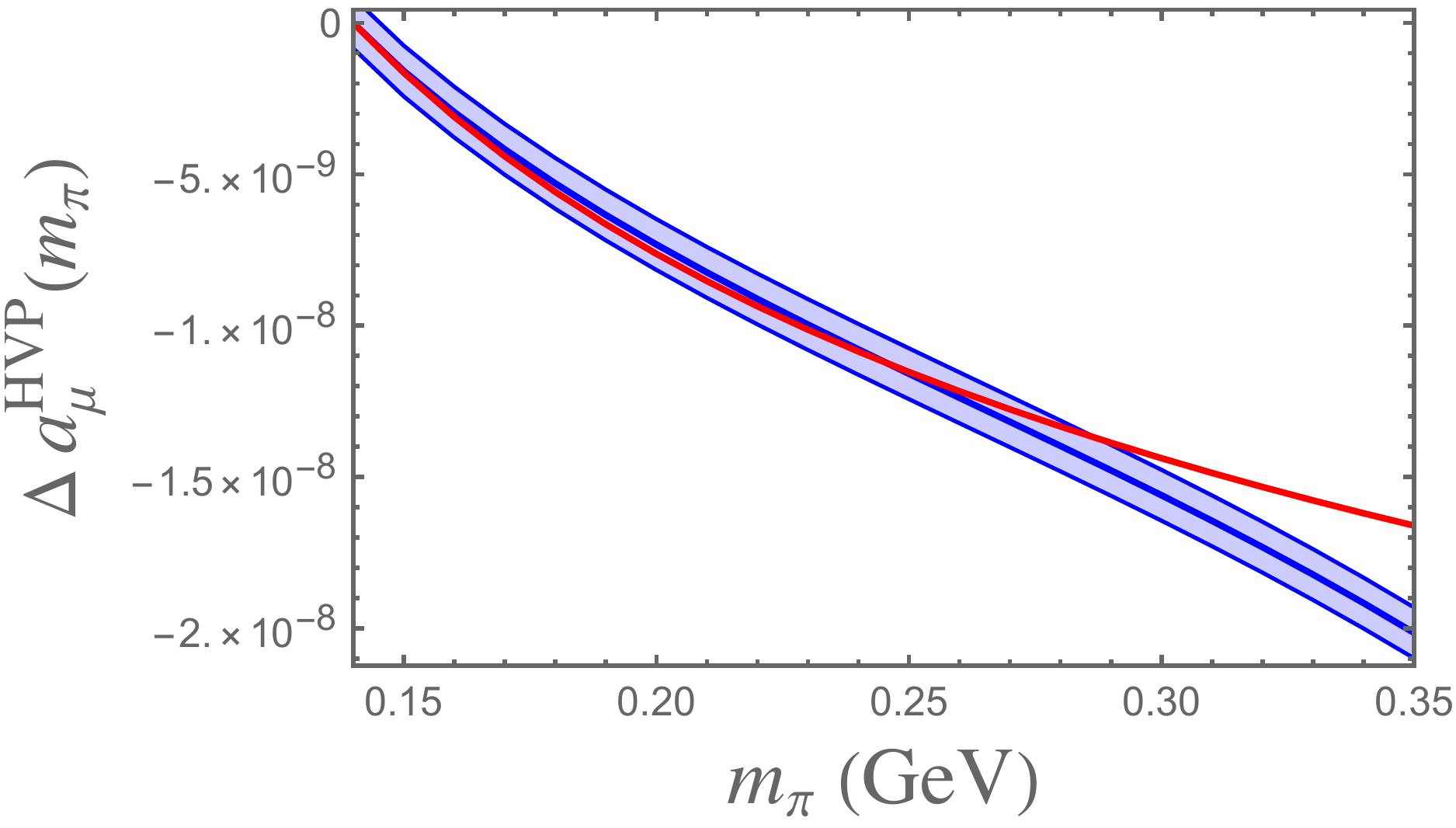}
\end{center}
\begin{quotation}
\caption{{\it Comparison of $\Delta a_\mu^{\rm HVP}(m_\pi)=a_\mu^{\rm HVP}(m_\pi)-a_\mu^{\rm HVP}(m_\pi=140\ \mbox{\rm MeV})$ between NNLO ChPT, and the resummation of Ref.~\cite{Coletal}, as a function of $m_\pi$.
The blue band gives the result of Ref.~\cite{Coletal}; the red curve is computed with NNLO ChPT.
\label{fig1}}}
\end{quotation}
\vspace*{-10ex}
\end{figure}
We can also use ChPT in order to estimate the dependence of $a_\mu^{\rm HVP}$ on $m_\pi$, 
as long as $m_\pi$ is not too large.   In Fig.~\ref{fig1} we show the change $\Delta a_\mu^{\rm HVP}$
with the pion mass.   The red curve gives the result of NNLO ChPT, while the blue band is the 
prediction for $\Delta a_\mu^{\rm HVP}$ from Ref.~\cite{Coletal}, which resums NNLO ChPT using
the Omn\`es relation and the inverse amplitude method, and which agrees well with experiment
(the band indicates their error estimate).   We see that NNLO ChPT (without any resummation)
agrees very well up to a pion mass of about 280~MeV.   Note that the LEC $c_{56}$ drops out of
this difference, thus making the NNLO ChPT prediction for $\Delta a_\mu^{\rm HVP}$ much more precise
than for $a_\mu^{\rm HVP}$ itself.   

\begin{figure}[t!]
\vspace*{4ex}
\begin{center}
\includegraphics*[width=7cm]{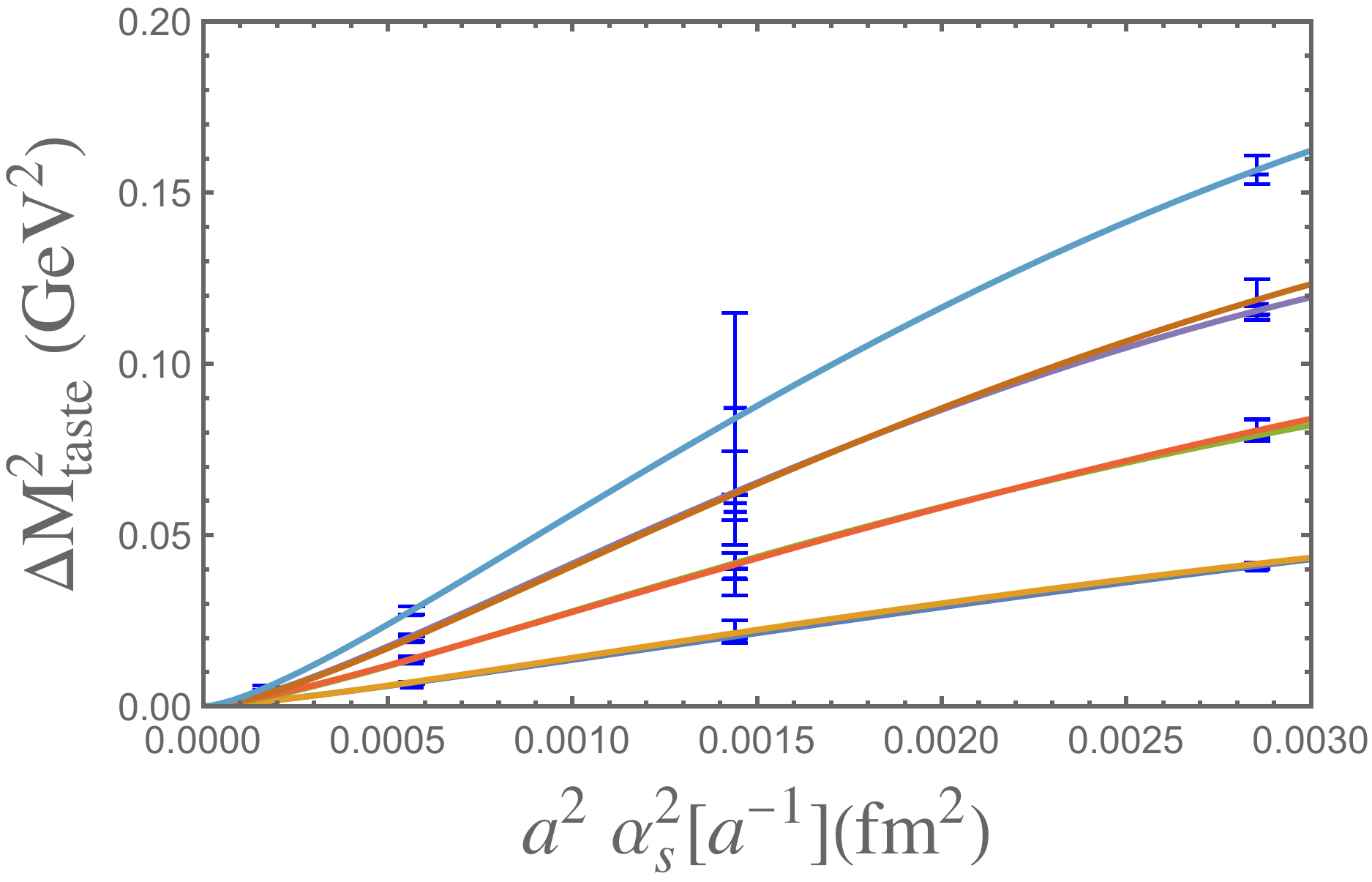}
\hspace{0.5cm}
\includegraphics*[width=7cm]{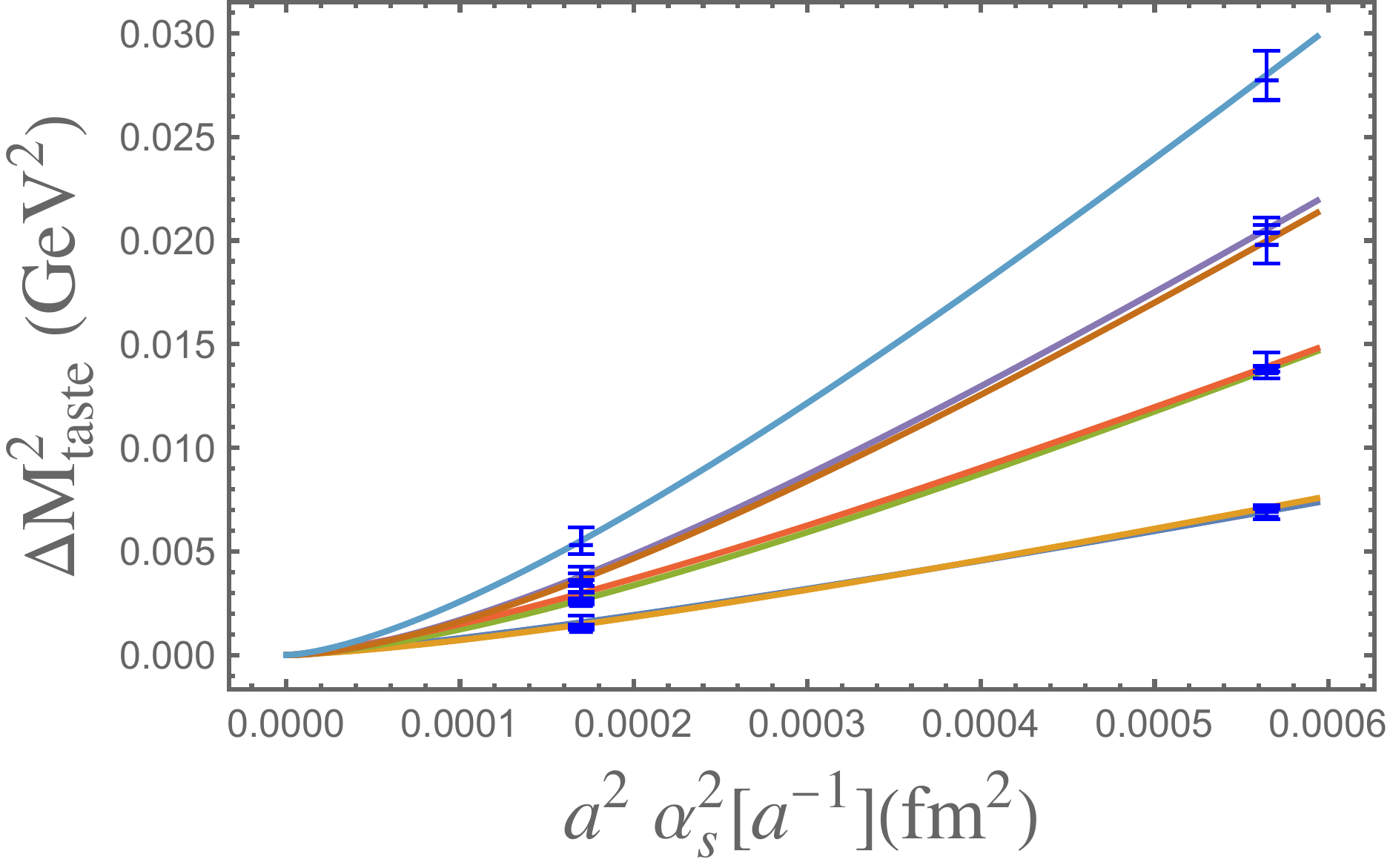}
\end{center}
\begin{quotation}
\caption
{{\it Pion taste splittings as a function of $a^2\alpha_s^2(1/a)$, with
$\alpha_s(1/a)$ the $\overline{{\rm MS}}$ coupling at scale $1/a$.  Points with error bars correspond to the measured taste splittings on the four MILC ensembles with lattice spacings $0.057$, $0.088$, $0.12$ and $0.15$~fm. The curves reflect fits to the form $m^2_X-m^2_\pi=A_Xa^2\alpha^2_s(1/a)+B_X a^4+C_X a^6$,
where $X$ labels the tastes $k$, $4$, $jk$, $k4$, $k5$ and $45$ (see Ref.~\cite{Aubin:2022hgm}).  The figure on the right zooms in on the region with the two smaller lattice spacings, $a=0.057$~fm and $a=0.088$~fm.  The near-degeneracy of the $X=k$ and $X=4$, $X=jk$ and $X=k4$, and $X=k5$ and $X=45$ taste
splittings is clearly visible.\label{tastesplittings}}}
\end{quotation}
\vspace*{-2ex}
\end{figure}
We conclude from this comparison that NNLO staggered ChPT (SChPT) can be used to correct for
taste-breaking corrections (as well as finite-volume effects and pion mass retuning), as long as the maximum pion mass in a taste multiplet is not larger than
about 300~MeV.   The taste splittings for our ensembles as a function of the lattice spacing are shown in Fig.~\ref{tastesplittings}, along with an NNLO-SChPT-inspired fit.   The smallest pion mass for each
ensemble is physical, while the largest pion masses are 153, 212, 326 and 418~MeV increasing with
the lattice spacing.   With the latter two masses being larger than 280~MeV, the application of NNLO 
SChPT to the coarser ensembles is questionable.   This is consistent with the empirical observation
that terms up to order $a^6$ are needed to fit the taste splittings, as shown in Fig.~\ref{tastesplittings}.
In fact, there is a puzzle: while SChPT predicts a leading-order behavior of the taste splittings that 
goes like $a^2$, the coefficients $A_X$ in the fit turn out to be consistent with zero.   Nevertheless, the
approximate degeneracies seen in the taste spectrum are a prediction of leading-order SChPT
\cite{LS}.   It is clear, however, that the behavior of the taste splittings is quite non-linear in $a^2$,
which suggests that at least the coarser ensembles are not reliably within reach of SChPT.\\

\section{Results and discussion}
\begin{figure}[t!]
\vspace*{4ex}
\begin{center}
\includegraphics*[width=15cm]{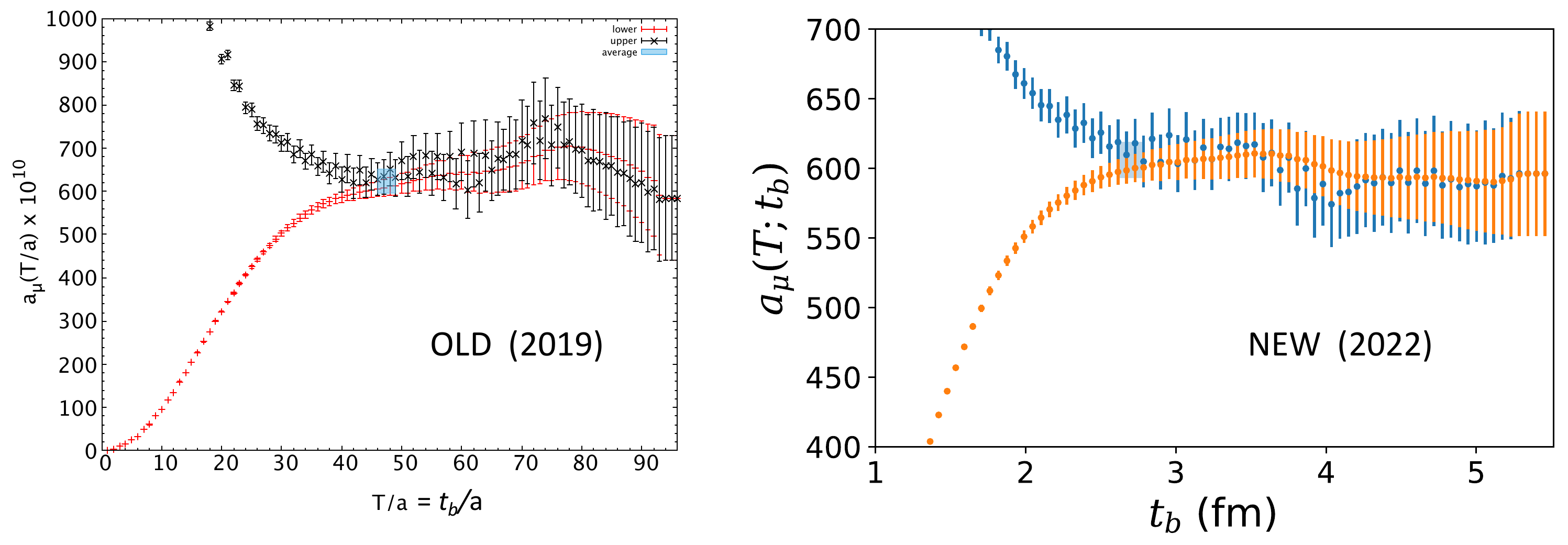}
\end{center}
\vspace*{-6ex}
\begin{quotation}
\caption{{\it $a_\mu^{\rm HVP,lqc}$ as a function of the bounding transition
$t_b$ on the 96 ensemble; from Ref.~\cite{ABGP19}
(left panel), and from Ref.~\cite{Aubin:2022hgm} (right panel).   Notice the difference in vertical scale.
\label{fig3}}}
\end{quotation}
\vspace*{-6ex}
\end{figure}

\begin{table}[t]
\begin{center}
\begin{tabular}{|c|c|c|c|c|c|c|c|c|} 
\hline
label & $a$ (fm) & $L^3\times T$ & $m_\pi$ (MeV) & $m_S$ (MeV) & $m_\pi L$ & \#configs & sep. & \#low modes \\
\hline
96 & 0.05684 & $96^3\times 192$ & 134.3 & 153 & 3.71 & 77 & 60 & 8000\\
64 & 0.08787 & $64^3\times 96$ & 129.5 & 212 & 3.69 & 78 & 100 & 8000\\
48I & 0.12121 & $48^3\times 64$ & 132.7 & 326 & 3.91 & 32 & 100 & 8000\\
32 & 0.15148 & $32^3\times 48$ & 133.0 & 418 & 3.27 & 48 & 40 & 8000 \\
48II & 0.15099 & $48^3\times 64$ & 134.3 & 418 & 4.93 & 40 & 100 & 8000\\
\hline
\end{tabular}
\end{center}
\vspace*{-3ex}
\begin{quotation}
\caption{{\it Parameters defining the lattice ensembles.   Columns contain a label to refer to the ensemble, the lattice spacing $a$, the spatial volume $L^3$ times the temporal direction $T$ (in lattice units), the Nambu--Goldstone pion mass $m_\pi$, the maximum pion mass $m_S$ in the pion taste multiplet, $m_\pi L$, the number of configurations in the ensemble, the separation between measurements ({\rm ``sep.''}), and the number of low-mode eigenvectors.
\label{tab1}}}
\end{quotation}
\vspace*{-8.5ex}
\end{table}
We now turn to the results we obtained in Ref.~\cite{Aubin:2022hgm}, to which we refer for details.   We computed the light-quark-connected part $a_\mu^{\rm HVP,lqc}$ on the ensembles shown in Table~\ref{tab1}.  On these ensembles, we have computed
\begin{equation}
\label{lqc}
a_\mu^{\rm HVP,lqc}= 2\alpha^2\sum_0^{T/2} w(t)\,C_{\rm lqc}(t)\ ,\qquad
C_{\rm lqc}(t)=\frac{1}{3}\sum_{\vec{x}}\sum_{i=1}^3\langle j^{\rm EM}_i(\vec{x},t)j^{\rm EM}_i(0)\rangle\ ,
\end{equation}
where $w(t)$ is a known weight function \cite{BM11}, and $C_{\rm lqc}(t)$ is the light-quark-connected 
correlation function of the conserved hadronic EM current $j_\mu^{\rm EM}$.   We used the bounding method of
Ref.~\cite{RBC18}, with a transition to the bounds around $t_b=3$~fm.
 The difference with our earlier results
\cite{ABGP19} is that we added two ensembles at a lattice spacing $0.15$~fm, and we added more 
configurations, better separation, and more low-mode eigenvectors.   This leads to more precise results,
as shown, for example, in Fig.~\ref{fig3} for the 96 ensemble.   

\begin{figure}[t!]
\vspace*{4ex}
\begin{center}
\includegraphics*[width=9cm]{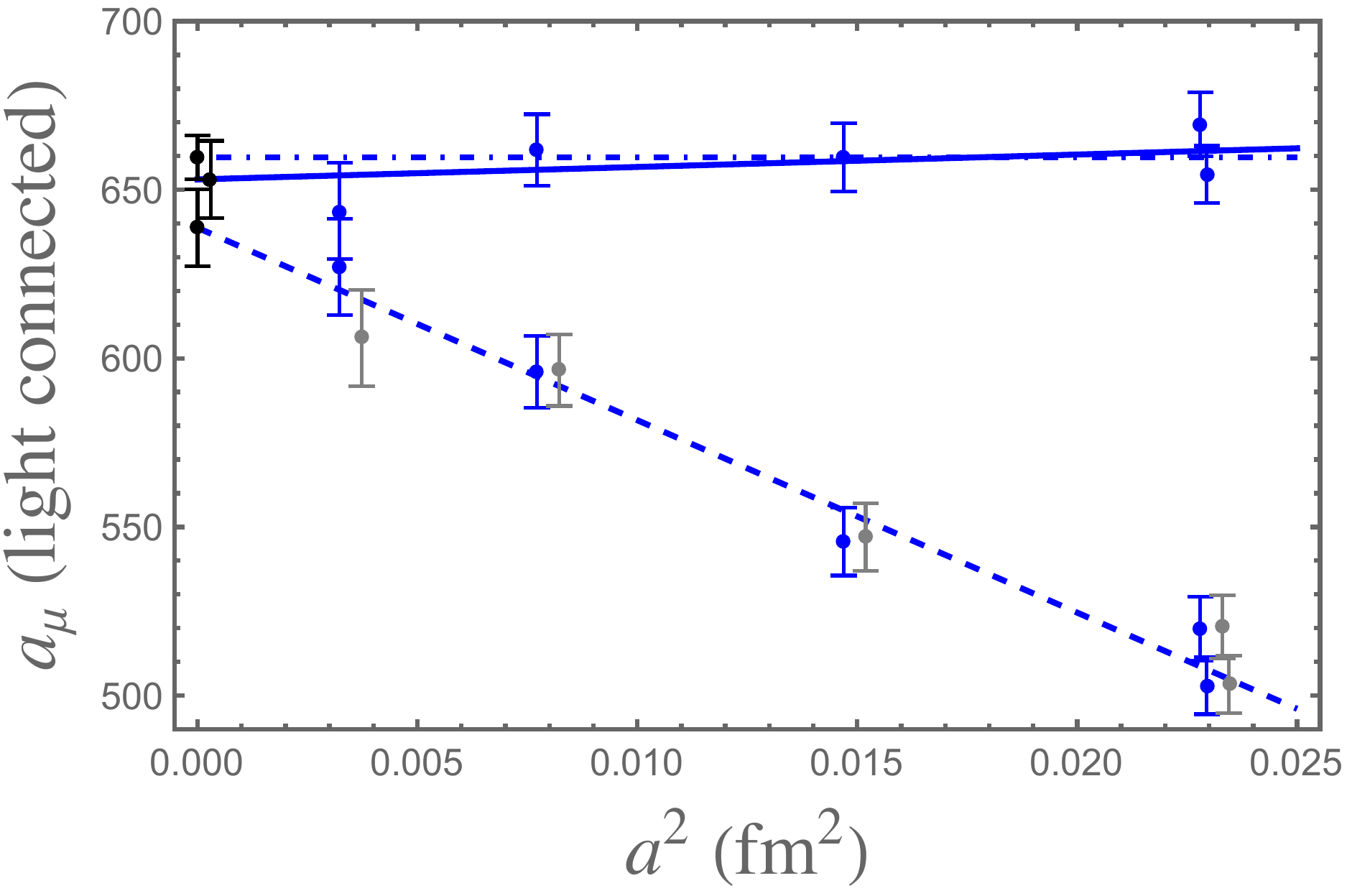}
\end{center}
\vspace*{-6ex}
\begin{quotation}
\caption%
{{\it $a_\mu^{\rm HVP,lqc}$ as a function of $a^2$, in units of $10^{-10}$. Fits using NNLO SChPT for FV corrections and pion mass retuning; without taste breaking  (linear, blue dashed line) and with taste breaking (linear, blue solid line or constant, blue dot-dashed line).      Continuum limits are shown in black.   The grey points in both panels (slightly horizontally offset for clarity) show the uncorrected values.
Some continuum-limit
extrapolations are slightly offset for clarity. \label{amufig}}}
\end{quotation}
\vspace*{-4ex}
\end{figure}
In Fig.~\ref{amufig} we show our results for $a_\mu^{\rm HVP,lqc}$ as a function of $a^2$.    The lower
blue data points are corrected for finite-volume (FV) effects and for pion-mass mistuning using NNLO
ChPT, whereas the higher blue data points are also corrected for taste breaking (the grey points are
uncorrected).  Solid and dashed lines are linear fits; the dot-dashed line is a constant fit.
\begin{table}[t]
\begin{center}
\begin{tabular}{|c|c|c|c|c|}
\hline
 & $a_\mu(96)-a_\mu(64)$ & $a_\mu(96)-a_\mu(\mbox{48I})$ & $a_\mu(96)-a_\mu(32)$ & $a_\mu(96)-a_\mu(\mbox{48II})$\\
\hline
lattice & 10(16) & 59(16) & 103(15) & 86(15)\\
\hline
NLO~SChPT & 11 & 28 & 38 & 37\\
NNLO~SChPT & 28 & 75 & 114 & 111 \\
\hline
SRHO & 35 & 89 & 129 & 128 \\
\hline
\end{tabular}
\end{center}
\vspace*{-3ex}
\begin{quotation}
\caption{{\it Differences of $a_\mu^{\rm HVP,lqc}$ values between different ensembles.
All number in units of $10^{-10}$; $a_\mu\equiv a_\mu^{\rm HVP,lqc}$. \label{tab:diffs}}}
\end{quotation}
\vspace*{-4.5ex}
\end{table}
Following Ref.~\cite{BMW20}, to test ChPT, we show in Table~\ref{tab:diffs} the differences between $a_\mu^{\rm HVP,lqc}$ values
computed on different ensembles, comparing both SChPT and the SRHO model of Ref.~\cite{HPQCD16}
with the data.   This table shows that, albeit within rather large errors, both NNLO SChPT (an EFT) and SRHO
(a model) describe the data fairly well, while NLO SChPT is not sufficient.   From Fig.~\ref{amufig} and the discussion of taste breaking in the
previous section, we conclude that smaller lattice spacings will probably be needed.   For instance, if we
would extrapolate from the smallest two lattice spacings, the continuum-limit value of $a_\mu^{\rm HVP,lqc}$ would come out smaller than the values shown in the figure, at the very low end of the result we obtained
in Ref.~\cite{Aubin:2022hgm}, which is $646(14)\times 10^{-10}$.   Likewise, better statistics and a more
precise scale setting will
be needed in order to reduce the error much below the 2.2\% relative error of this result.

\begin{figure}[t]
\vspace*{4ex}
\begin{center}
\includegraphics*[width=7cm]{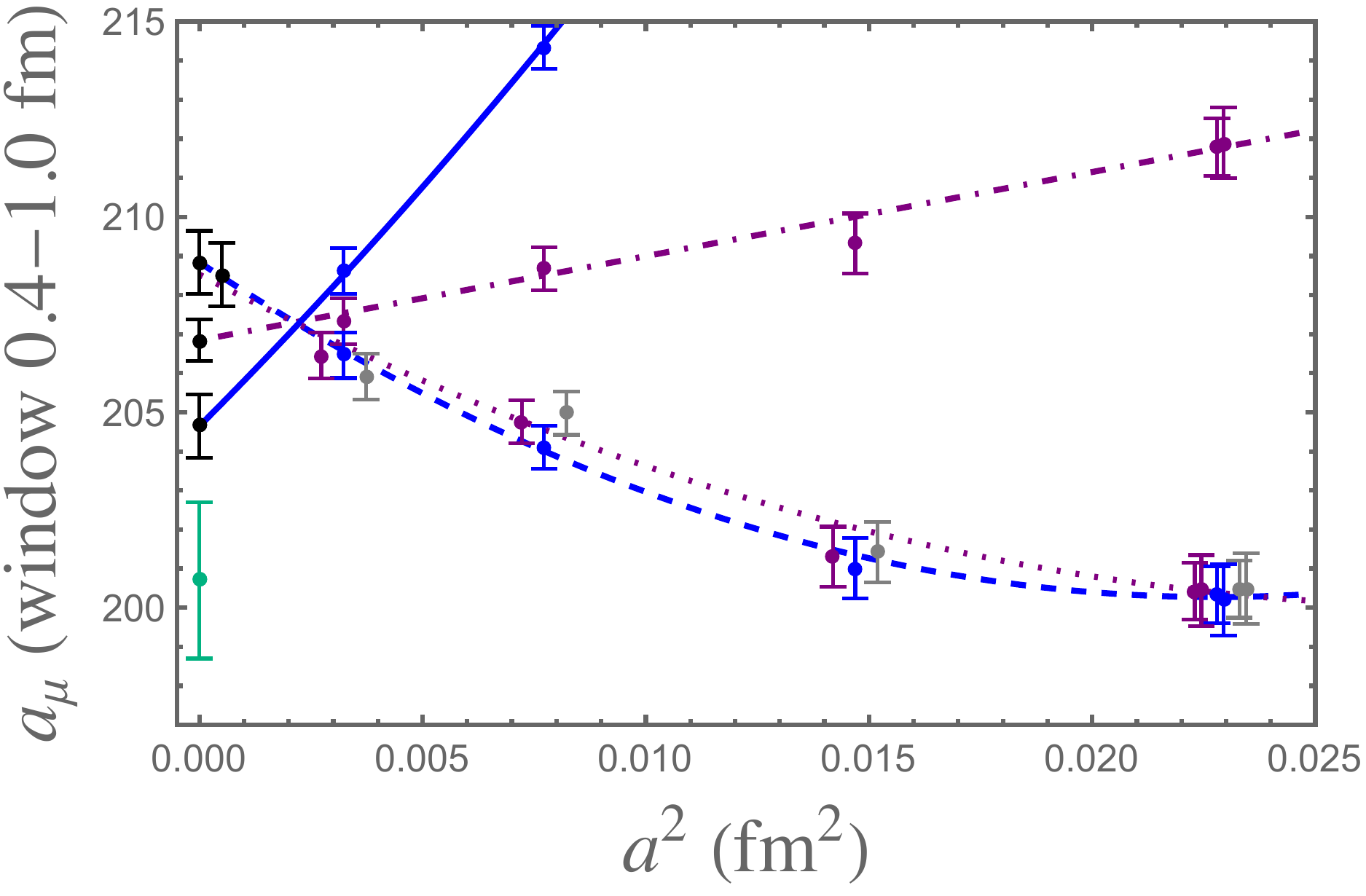}
\hspace{0.5cm}
\includegraphics*[width=7cm]{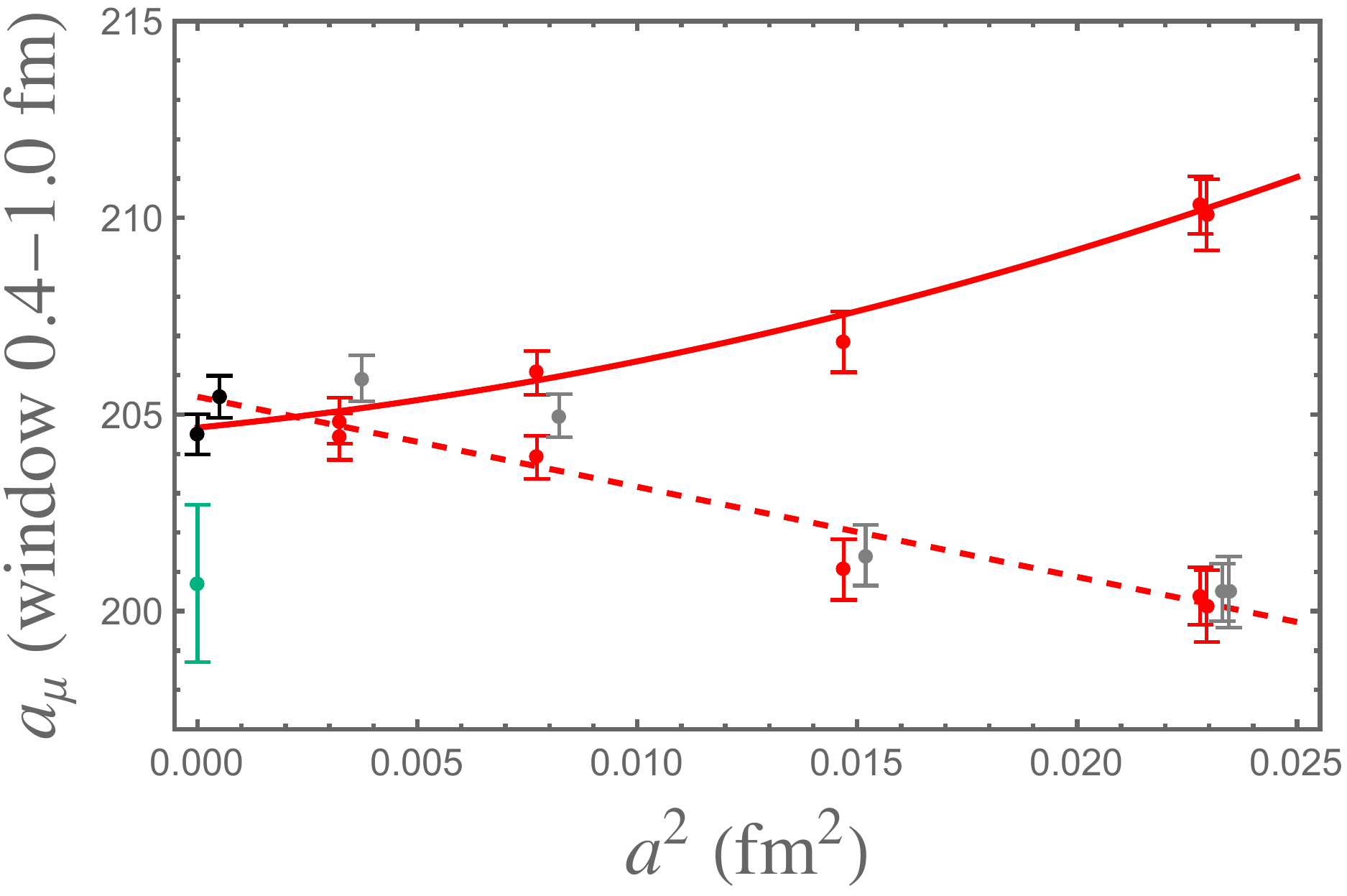}
\end{center}
\begin{quotation}
\caption%
{{\it The 0.4-1.0~{\rm fm} window, $a_\mu^{\rm W1,lqc}$, in units of $10^{-10}$. Left panel:  Fits using NNLO SChPT for FV corrections and pion mass retuning; without taste breaking  (quadratic, blue dashed line) and with taste breaking (quadratic, blue solid curve); fits using NLO SChPT
for FV corrections and pion mass retuning; without taste breaking  (quadratic, purple dotted line) and with taste breaking (linear, purple dot-dashed line).   Right panel:
Fits using the SRHO model for FV corrections and pion mass retuning; without taste breaking  (linear, red dashed line) and with taste breaking (quadratic, red solid curve).   The data points for each fit are shown in the same color; continuum limits are shown in black.   The grey points in both panels (slightly horizontally offset for clarity) show the uncorrected values.
The isolated (green) point at $a^2=0$ is the estimated value from R-ratio data (by C.~Lehner, using data from 
Ref.~\cite{KNT18}). Some data points and continuum-limit
extrapolations are slightly offset for clarity. \label{window1}}}
\end{quotation}
\vspace*{-4ex}
\end{figure}
The ``standard intermediate window" between $t=0.4$ and $1.0$~fm of Ref.~\cite{RBC18} can be computed much more precisely, allowing us to scrutinize these issues in more detail with the same lattice data, as this 
window is centered on
the region where statistical and systematic errors are smaller than for  $a_\mu^{\rm HVP,lqc}$ itself.
An important caveat is that effective field theory methods are not applicable to this intermediate-distance
quantity, as discussed in more detail in Ref.~\cite{Aubin:2022hgm}, so all approaches to correcting for
systematic effects, including ChPT, should be considered model dependent.   Our results for this window
(W1) are shown in Fig.~\ref{window1}.
A table similar to 
Table~\ref{tab:diffs} \cite{Aubin:2022hgm} (not shown here) shows that both NLO SChPT and the SRHO model describe
lattice data reasonably well, while NNLO SChPT is completely off, as also suggested by the NNLO
curve in Fig.~\ref{window1}.   This figure shows that there is a significant curvature as a function of
$a^2$, while continuum extrapolations of data uncorrected or corrected for taste breaking do not
agree.   This again strongly suggests that smaller lattice spacings will be needed in order to control
the continuum limit.   Our value for $a_\mu^{\rm W1,lqc}$ is $206.8(2.2)\times 10^{-10}$, with a 
relative error about half that for $a_\mu^{\rm HVP,lqc}$.   However, in this case, the combined error
is dominated by systematics; the statistical error being about a third of the total error.   We note that
the right panel of Fig.~\ref{window1} is very similar to Fig.~4 in Ref.~\cite{BMW20}, suggesting that
both their and our results are approximately equally far from the continuum limit.

\begin{figure}[t]
\vspace*{4ex}
\begin{center}
\includegraphics*[width=10cm]{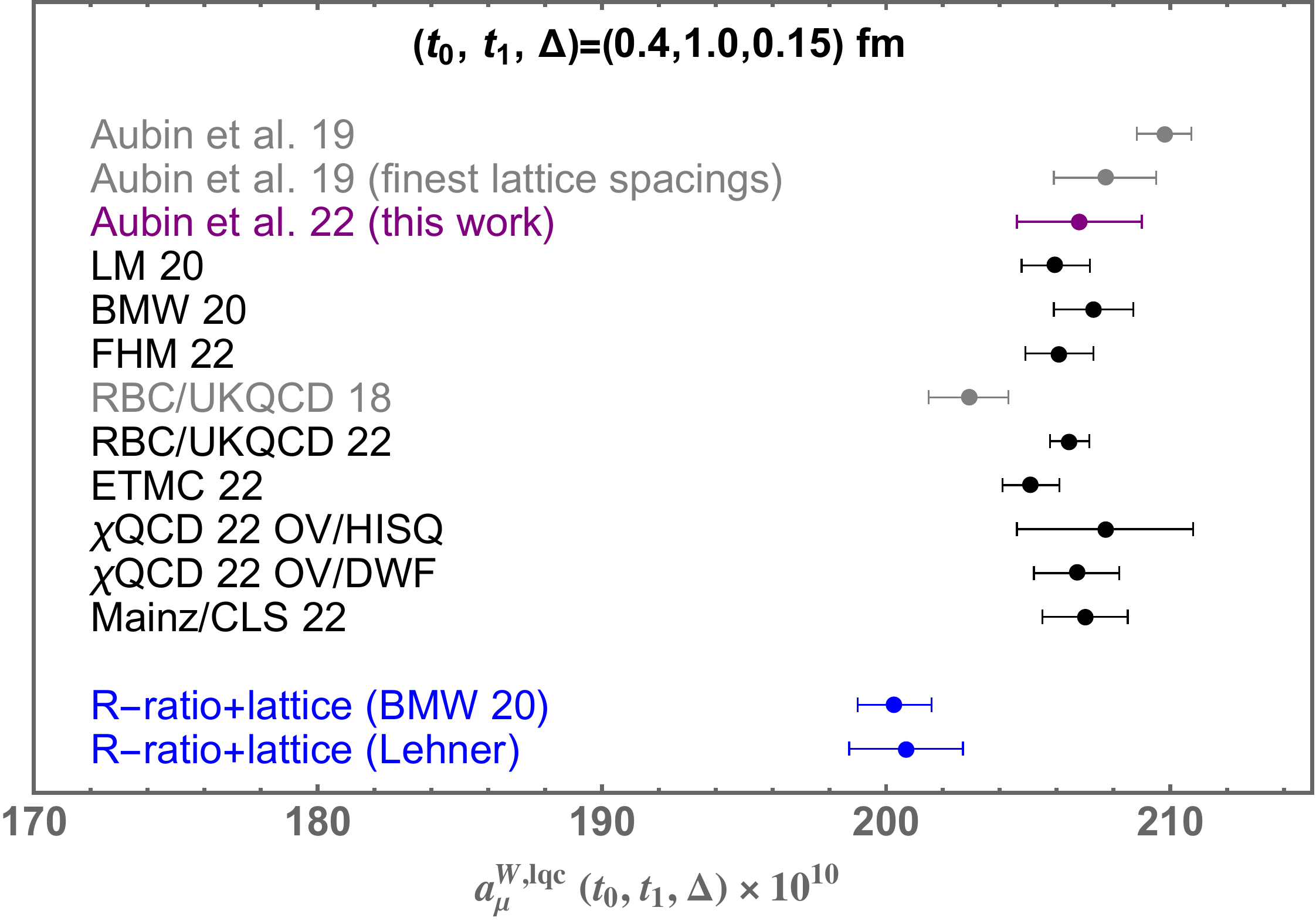}
\end{center}
\begin{quotation}
\caption%
{{\it Recent values of $a_\mu^{\rm W1,lqc}$, compared to values obtained from the $R$-ratio data of
Ref.~\cite{KNT18} subtracting non-light-quark-connected contributions computed on the lattice.  This
work's value is shown in purple, while some superseded values are shown in grey. 
The references are ABGP~{\it et al.}~19 \cite{ABGP19}, ABGP~{\it et al.}~22 \cite{Aubin:2022hgm},
LM~20 \cite{LM20}, BMW~20 \cite{BMW20}, 
FHM~22 \cite{FHM22}, RBC/UKQCD~18 \cite{RBC18}, RBC/UKQCD~22 \cite{RBC22},
ETMC~22 \cite{ETMC22}, $\chi$QCD~22~OV/HISQ and $\chi$QCD~22~OV/DWF \cite{chiQCD22}
and Mainz~22 \cite{Mainz22}.
\label{summary}}}
\end{quotation}
\vspace*{-4ex}
\end{figure}
Figure~\ref{summary} summarizes recent values of the window quantity $a_\mu^{\rm W1,lqc}$
obtained in the literature.   While we believe that the understanding of systematic errors can and
should be improved, it is intriguing that all lattice-based values appear to agree with each other.
We refrain from presenting a ``lattice average'' here, but it is clear that such an average would be
significantly larger than the $R$-ratio based value.   

\begin{figure}[t]
\vspace*{4ex}
\begin{center}
\includegraphics*[width=7cm]{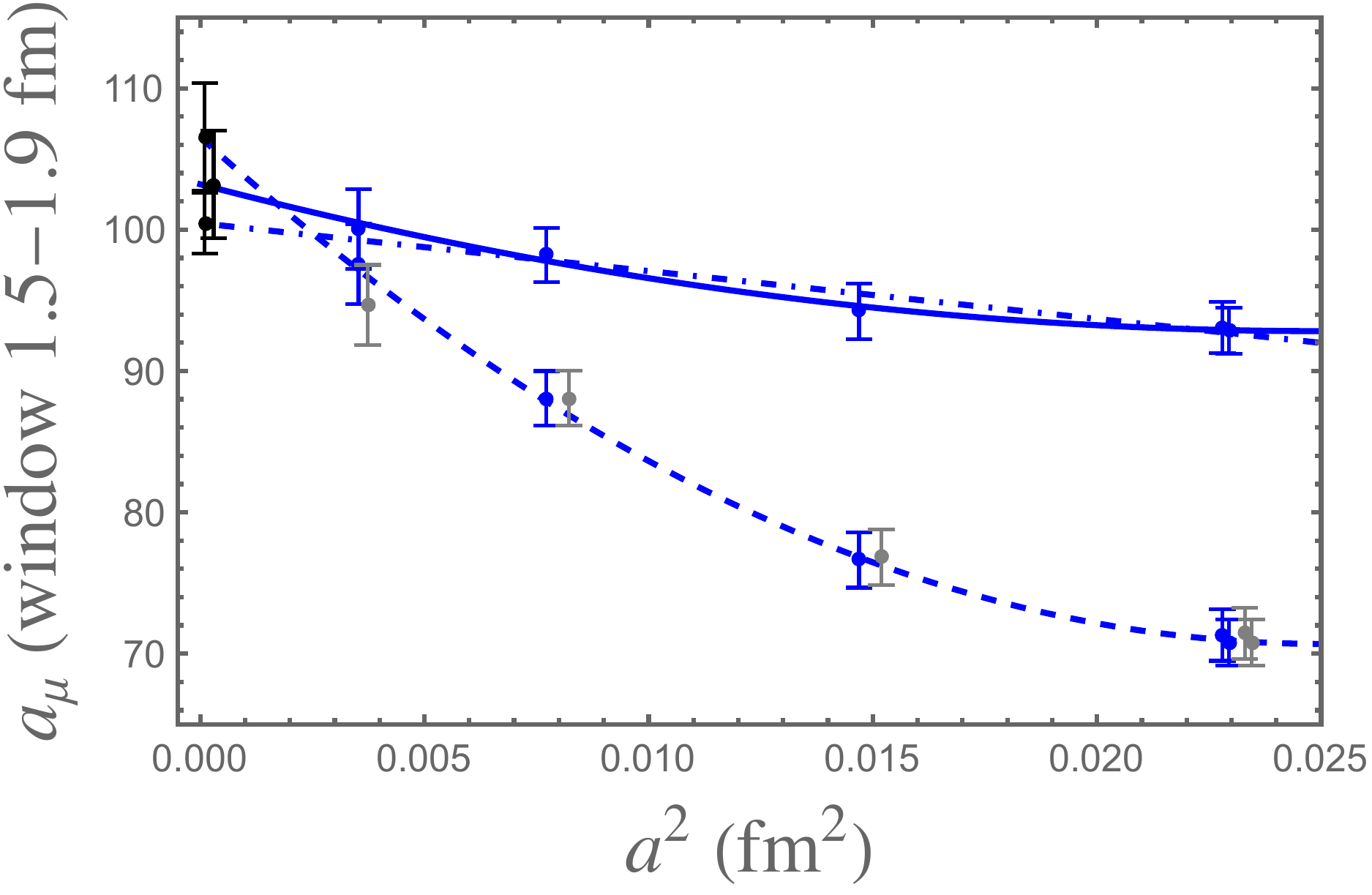}
\hspace{0.5cm}
\includegraphics*[width=7cm]{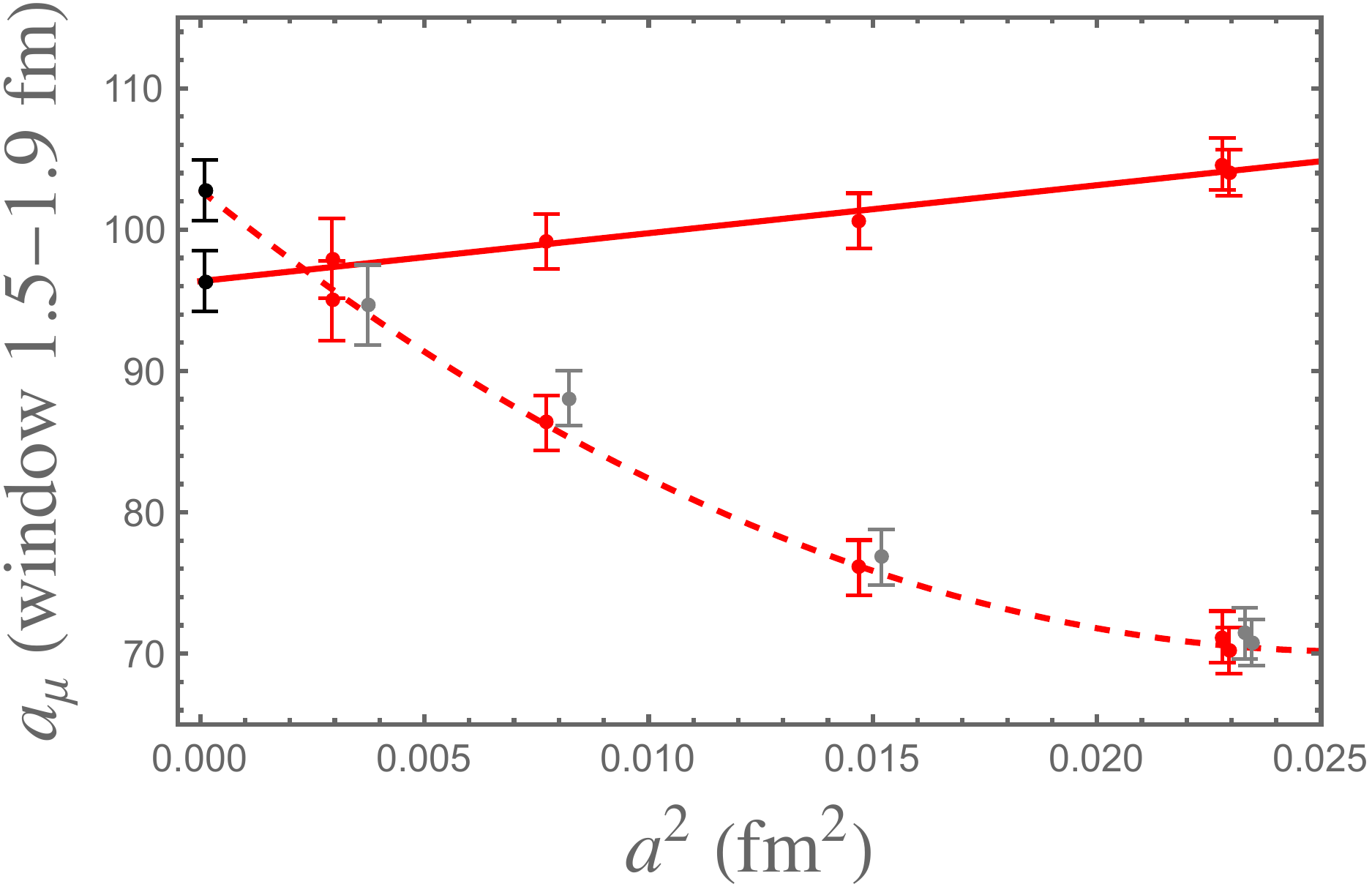}
\end{center}
\vspace*{-4ex}
\begin{quotation}
\caption%
{{\it The 1.5-1.9~{\rm fm} window, $a_\mu^{\rm W2,lqc}$, in units of $10^{-10}$. Left panel:  Fits using NNLO SChPT for FV corrections and pion mass retuning; without taste breaking  (quadratic, blue dashed line) and with taste breaking (quadratic, blue solid curve; linear, dot-dashed line).   Right panel:
Fits using the SRHO model for FV corrections and pion mass retuning; without taste breaking  (linear, red dashed line) and with taste breaking (quadratic, red solid curve).   The data points for each fit are shown in the same color; continuum limits are shown in black.   The grey points in both panels (slightly horizontally offset for clarity) show uncorrected values. Some data points and continuum-limit
extrapolations are slightly offset for clarity. \label{window2}}}
\end{quotation}
\vspace*{-4ex}
\end{figure}
\begin{table}[t]
\begin{center}
\begin{tabular}{|c|c|c|c|c|}
\hline
 & ${\rm W2}(96)-{\rm W2}(64)$ & ${\rm W2}(96)-{\rm W2}(48\mbox{I})$ & ${\rm W2}(96)-{\rm W2}(32)$ & ${\rm W2}(96)-{\rm W2}(48\mbox{II})$ \\
\hline
lattice & 6.6(2.7) & 17.8(2.8) & 23.9(2.6) & 23.2(2.7) \\
\hline
NLO~SChPT & 2.1 & 5.0 & 6.7 & 6.4 \\
NNLO~SChPT & 4.7 & 12.0 & 16.7 & 16.3  \\
\hline
SRHO & 7.8 & 20.5 & 30.0 & 29.9 \\
\hline
\end{tabular}
\end{center}
\vspace*{-3ex}
\begin{quotation}
\caption{{\it Differences of $a_\mu^{\rm W2,lqc}$ values between different ensembles.
All number in units of $10^{-10}$; ${\rm W2}\equiv a_\mu^{\rm W2,lqc}$. \label{tab:diffsW2}}}
\end{quotation}
\vspace*{-4.5ex}
\end{table}
Finally, we proposed a new window quantity, with a window W2 between 1.5 and 1.9~fm (with the same
smoothing as the W1 window, $\Delta=0.15$~fm).   The motivation is that this window is a longer distance
window, which should be accessible to ChPT, while still easier to compute on the lattice than the full
$a_\mu^{\rm HVP}$.   We show results for $a_\mu^{\rm W2,lqc}$ in Fig.~\ref{window2}, and we obtained the
continuum-limit value $102.1(2.4)\times 10^{-10}$ in Ref.~\cite{Aubin:2022hgm}, with equal values from linear  fits to NNLO SChPT with and without taste-breaking corrections if we drop the data points at the largest lattice spacing.    The error is indeed
much smaller than the error on $a_\mu^{\rm HVP,lqc}$, and about the same as the error on
$a_\mu^{\rm W1,lqc}$; 
differences
analogous to those of Table~\ref{tab:diffs} are shown in Table~\ref{tab:diffsW2}.   We see from this table that
indeed NNLO SChPT does reasonably well in describing the data, although it underestimates the lattice
differences about as much as the SRHO model overestimates them.     The consistency of our fits
without the largest lattice spacing, together with the curvature of most fits in Fig.~\ref{window2},
suggests again that smaller lattice spacings will be needed, with better statistics, in order to obtain a 
reliable continuum limit with a smaller error. 

\section{Conclusion}
We briefly summarize our conclusions, referring to Ref.~\cite{Aubin:2022hgm} for many of the details:
\begin{itemize}
\item ChPT {\it can} be applied to $a_\mu^{\rm HVP}$, if the pion masses (including taste partners in the
staggered case) are small enough.  ChPT does {\it not} work for short or intermediate distance windows,
for which no systematic EFT approach is available.   Since, at present, systematic errors in such windows
are still not negligible, models are needed for the comparison between different lattice discretizations.
\item Staggered computations need lattice spacings significantly smaller than 0.06~fm to control the continuum limit;  at present, staggered computations appear to be far from the linear $a^2$ regime.
Related, a better understanding of taste splittings would be desirable.
\item We propose to consider longer-distance windows, such as the window W2 considered here and
in Ref.~\cite{Aubin:2022hgm}.   Longer-distance windows are accessible to ChPT, while they still can be
computed with higher statistical precision than $a_\mu^{\rm HVP}$ itself.
\item Finally, scaling setting errors in our computation are of the same order as statistical errors.   
Improvement of scale setting would thus also help to reduce overall errors.
\end{itemize}

\vspace{3ex}
\noindent {\bf Acknowledgments}
\vspace{3ex}

We thank Claude Bernard and Kim Maltman for discussions,
Doug Toussaint for providing the full taste-pion spectrum on the
96, 64, 48I and 32 ensembles, the MILC collaboration for the use of their configurations,
and Andr\'e Walker-Loud (for CalLat) for ensemble 48II.
This work used the Extreme Science and Engineering Discovery Environment (XSEDE),
which is supported by National Science Foundation grant number ACI-1548562.
We thank the Pittsburgh Supercomputing Center
(PSC), the San Diego Supercomputer Center (SDSC), and  the Texas Advanced Computing
Center (TACC), 
where the lattice computations were performed. 
This material is based upon work supported by the U.S. Department 
of Energy, Office of Science, Office of Basic Energy Sciences Energy 
Frontier Research Centers program under Award Numbers DE-SC-0010339 (TB) and DE-SC-0013682 (MG).
SP is supported 
by the Spanish Ministry of Science, Innovation and Universities 
(project PID2020-112965GB-I00/AEI/10.13039/501100011033) and by 
Grant 2017 SGR 1069. IFAE is partially funded by the CERCA 
program of the Generalitat de Catalunya.


\begin{thebibliography}{99}

\bibitem{Aubin:2022hgm}
C.~Aubin, T.~Blum, M.~Golterman and S.~Peris,
Phys. Rev. D \textbf{106}, no.5, 054503 (2022)
[arXiv:2204.12256 [hep-lat]].

\bibitem{Aubin:2020scy}
C.~Aubin, T.~Blum, M.~Golterman and S.~Peris,
Phys. Rev. D \textbf{102}, no.9, 094511 (2020)
[arXiv:2008.03809 [hep-lat]].

\bibitem{RBC18}
T.~Blum \textit{et al.} [RBC and UKQCD],
Phys. Rev. Lett. \textbf{121}, no.2, 022003 (2018)
[arXiv:1801.07224 [hep-lat]].

\bibitem{Golterman:2017njs}
M.~Golterman, K.~Maltman and S.~Peris,
Phys. Rev. D \textbf{95}, no.7, 074509 (2017)
[arXiv:1701.08685 [hep-lat]].

\bibitem{MILC:2012znn}
A.~Bazavov \textit{et al.} [MILC],
Phys. Rev. D \textbf{87}, no.5, 054505 (2013)
[arXiv:1212.4768 [hep-lat]].

\bibitem{Miller:2020xhy}
N.~Miller, H.~Monge-Camacho, C.~C.~Chang, B.~H\"orz, E.~Rinaldi, D.~Howarth, E.~Berkowitz, D.~A.~Brantley, A.~S.~Gambhir and C.~K\"orber, \textit{et al.}
Phys. Rev. D \textbf{102}, no.3, 034507 (2020)
[arXiv:2005.04795 [hep-lat]].

\bibitem{LR}
B.~e.~Lautrup, A.~Peterman and E.~de Rafael,
Phys. Rept. \textbf{3}, 193-259 (1972)

\bibitem{TB}
T.~Blum,
Phys. Rev. Lett. \textbf{91}, 052001 (2003)
[arXiv:hep-lat/0212018 [hep-lat]].

\bibitem{Coletal}
G.~Colangelo, M.~Hoferichter, B.~Kubis, M.~Niehus and J.~R.~de Elvira,
[arXiv:2110.05493 [hep-ph]].

\bibitem{LS}
W.~J.~Lee and S.~R.~Sharpe,
Phys. Rev. D \textbf{60}, 114503 (1999)
[arXiv:hep-lat/9905023 [hep-lat]].

\bibitem{BM11}
D.~Bernecker and H.~B.~Meyer,
Eur. Phys. J. A \textbf{47}, 148 (2011)
[arXiv:1107.4388 [hep-lat]].

\bibitem{ABGP19}
C.~Aubin, T.~Blum, C.~Tu, M.~Golterman, C.~Jung and S.~Peris,
Phys. Rev. D \textbf{101}, no.1, 014503 (2020)
[arXiv:1905.09307 [hep-lat]].

\bibitem{HPQCD16}
B.~Chakraborty, C.~T.~H.~Davies, P.~G.~de Oliviera, J.~Koponen, G.~P.~Lepage and R.~S.~Van de Water,
Phys. Rev. D \textbf{96}, no.3, 034516 (2017)
[arXiv:1601.03071 [hep-lat]].

\bibitem{BMW20}
S.~Borsanyi, Z.~Fodor, J.~N.~Guenther, C.~Hoelbling, S.~D.~Katz, L.~Lellouch, T.~Lippert, K.~Miura, L.~Parato and K.~K.~Szabo, \textit{et al.}
Nature \textbf{593}, no.7857, 51-55 (2021)
[arXiv:2002.12347 [hep-lat]].

\bibitem{KNT18}
A.~Keshavarzi, D.~Nomura and T.~Teubner,
Phys. Rev. D \textbf{97}, no.11, 114025 (2018)
[arXiv:1802.02995 [hep-ph]].

\bibitem{LM20}
C.~Lehner and A.~S.~Meyer,
Phys. Rev. D \textbf{101}, 074515 (2020)
[arXiv:2003.04177 [hep-lat]].

\bibitem{FHM22}
S.~Gottlieb, talk at 
First LatticeNET workshop on challenges in Lattice field theory, Benasque, Sept. 2022,
https://www.benasque.org/2022lattice\_workshop/cgi-bin/talks/allprint.pl .

\bibitem{RBC22}
C.~Lehner, talk at Lattice 2022, Bonn, Aug. 2022,
https://indico.hiskp.uni-bonn.de/event/40/contributions/783/ .

\bibitem{ETMC22}
C.~Alexandrou, S.~Bacchio, P.~Dimopoulos, J.~Finkenrath, R.~Frezzotti, G.~Gagliardi, M.~Garofalo, K.~Hadjiyiannakou, B.~Kostrzewa and K.~Jansen, \textit{et al.}
[arXiv:2206.15084 [hep-lat]].

\bibitem{chiQCD22}
G.~Wang \textit{et al.} [chiQCD],
[arXiv:2204.01280 [hep-lat]].

\bibitem{Mainz22}
M.~C\`e, A.~G\'erardin, G.~von Hippel, R.~J.~Hudspith, S.~Kuberski, H.~B.~Meyer, K.~Miura, D.~Mohler, K.~Ottnad and P.~Srijit, \textit{et al.}
[arXiv:2206.06582 [hep-lat]].

\end{thebibliography}
\end{document}